\nonstopmode
\pdfoutput=1
\documentclass[apl,superscriptaddress,twocolumn,floatfix]{revtex4}
\usepackage{graphicx}
\usepackage{hyperref}
\usepackage{amsmath}
\usepackage[latin1]{inputenc} 
\usepackage[T1]{fontenc} 
\hypersetup{pdfauthor={some author},pdftitle={eye-catching title}}
\usepackage{natbib} 

\begin{document}
\title{$200\,ps$ Vortex Core Reversal by Azimuthal Spin Waves}
\author{Matthias Kammerer}
\author{Hermann Stoll}
\author{Matthias Noske}
\author{Markus Sproll}
\author{Markus Weigand}
\affiliation{Max Planck Institute for Intelligent Systems, Stuttgart, Germany\\ (formerly MPI for Metals Research)}
\author{Georg Woltersdorf}
\affiliation{Regensburg University, Regensburg, Germany}
\author{Gisela Schuetz}
\affiliation{Max Planck Institute for Intelligent Systems, Stuttgart, Germany\\ (formerly MPI for Metals Research)}
\date{\today}

\begin{abstract}

Spin wave mediated vortex core reversal has been investigated by time-resolved scanning transmission X-ray microscopy (STXM). Movies showing the development of the spin wave and vortex core magnetization dynamics during unidirectional vortex core reversal could be taken in Permalloy discs, $1.6\,\mu m$ in diameter and $50\, nm$-thick, during excitation with rotating ac field bursts of one period duration at $4.5\, GHz$ and with amplitudes up to $4\, mT$. Unidirectional switching is achieved by taking advantage of an asymmetry for CW or CCW excitation caused by the gyrofield. The differences in the magnetization dynamics due to this asymmetry could be imaged during continuous excitation with multi-$GHz$ rotating fields. All our experimental results are in good agreement with micromagnetic simulations. 

In addition, for the sample geometry given above, simulations reveal a lower limit of about $200\, ps$ for the time of unidirectional vortex core switching, which cannot be overcome by shortening the excitation length or by increasing the excitation amplitude. We explain this limitation by the finite time needed for an energy transfer of the global excitation towards the center of the sample. Thus smaller samples will allow for much shorter vortex core reversal times.

\end{abstract}

\maketitle

\begin{figure*}
  \centerline{\includegraphics[clip,width=\textwidth]{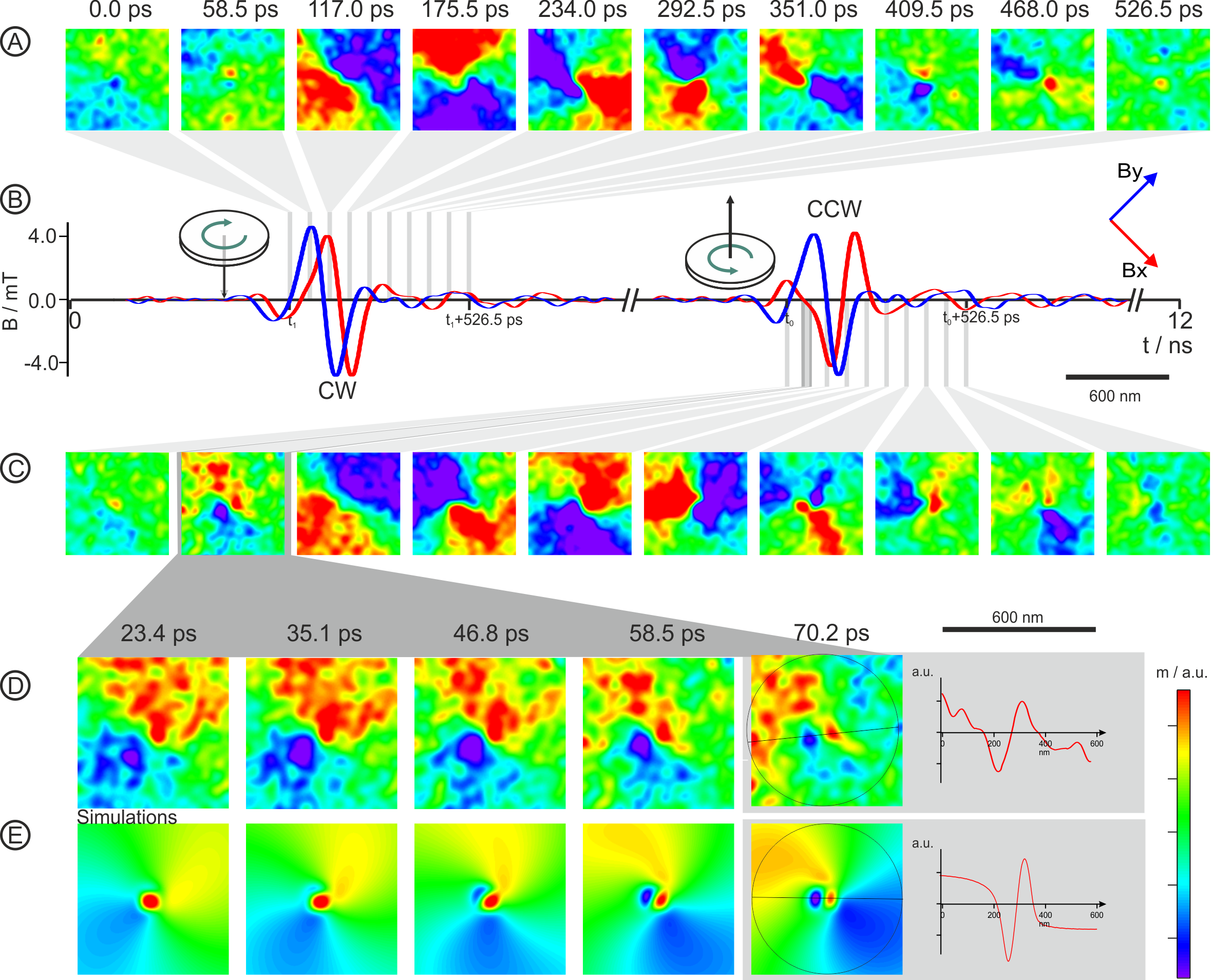}}
  \caption{{\bf Time-resolved imaging of spin wave mediated vortex core reversal.} B: Amplitudes in x and y direction of the CW (left) and CCW (right) rotating bursts of one period at $4.5\, GHz$. A and C: Snapshots in steps of $58.5\, ps$ showing the out-of-plane magnetization of the excited spin wave during unidirectional reversal of the vortex core polarization from down to up (A) and from up to down (B). D: Highly temporally resolved evolution of the spin wave amplitudes in time steps of $11.7\, ps$ starting from the beginning of the CCW excitation. The graph to the right shows a cut, as indicated by the line, through the most right image of the excitation mode (for details see text). E: Corresponding micromagnetic simulations, folded with the respective temporal and lateral resolution for a comparison with the experimental data (D).
  }
  \label{Fig:FastRotFieldsSwitching}
\end{figure*}

\begin{figure}
	\centerline{\includegraphics[clip,width=0.5\textwidth]{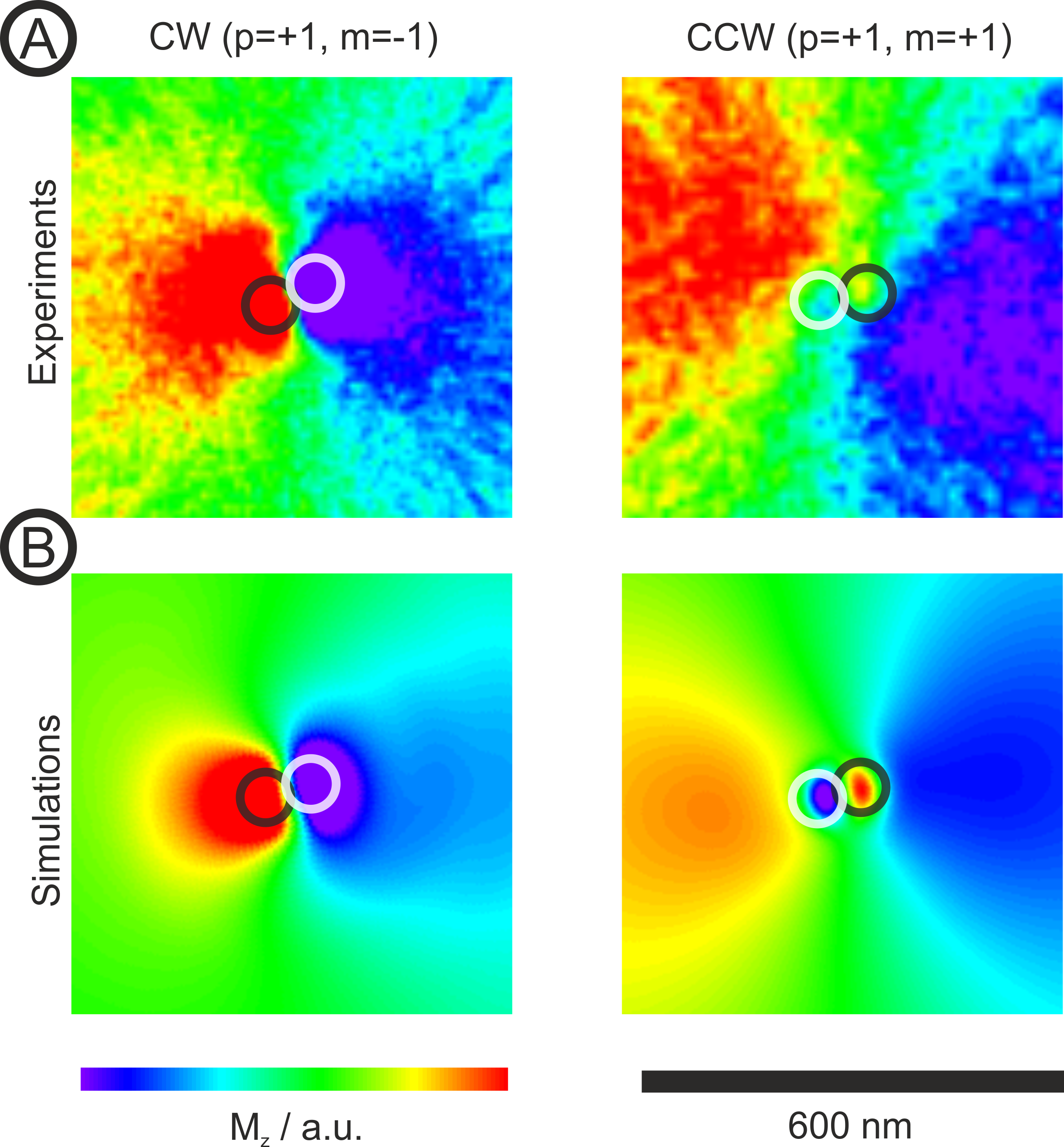}}
	\caption{{\bf Out-of-plane magnetization during continuous excitation just below the switching threshold for vortex core reversal.} 
	A: Experiment. B: Micromagnetic simulations. Left column: CW excitation of the azimuthal spin wave mode ($n=1, m=-1$). Right column: CCW excitation of the mode ($n=1, m=+1$). The difference in the magnetization dynamics between CW and CCW rotating spin waves are clearly visible in both, the experimental and the simulated data. For details see text. 	
	}
	\label{Fig:RotationSum}
\end{figure}

As fundamental spin structure of micron to nano-sized magnetic platelets the magnetic vortex \cite{Huber1982,Shinjo2000,Wachowiak2002} with an in-plane curling magnetization and a perpendicularly magnetized core has gained considerable attraction in the last years. Essential progress in the understanding of the nonlinear vortex dynamics was achieved by the experimental finding \cite{VanWaeyenberge2006}, that the vortex core, with its polarization pointing up ($p=+1$) or down ($p=-1$) can be reversed with low power by applying short bursts of alternating magnetic fields at $1.5\, mT$. In contrast to linear excitation which toggles the vortex core from its up to its down position and vice versa, unidirectional switching could be proofed experimentally by rotating magnetic fields \cite{Curcic2008,Curcic2011,Zagorodny2003,Lee2008,Kravchuk2007,Kim2008} and rotating currents \cite{Kamionka2010}. This low power and selective vortex core reversal \cite{Curcic2008} may open up the possibility of using the magnetic vortex core as memory bit \cite{Kim2008,Bohlens2008,Pigeau2010} and for further spintronic applications. So far, low-field selective vortex core reversal by gyrotropic mode excitation with frequencies in the sub-$GHz$ regime was limited to more than $1\, ns$. Beside this lowest excitation mode vortex structures possess azimuthal spin wave modes at high (multi-$GHz$) frequencies \cite{Park2003,Buess2004,Buess2005,Buess2005II,Guslienko2008101}, which are labeled in polar coordinates by a radial mode number $n$ and an azimuthal mode number $m$ with ($n>0,\,\|m\|>0$): 
\begin{equation}
\Delta  m_{z,n,m}(\rho, \phi,t)=  r_{z,n} (\rho) e^{i m \phi} e^{- i \omega_{n,m,p} t}
\end{equation}
The sign of $m$ denotes the rotation sense of the azimuthal wave (CCW for $m > 0$, CW for $m < 0$). 
The symmetry in the frequency splitting of the azimuthal mode can be written as $\omega_{n,m,-p} =\omega_{n,-m,p}$.

The interaction between the spin waves and the vortex core splits the frequency for CW and CCW rotating modes by lifting the degeneracy of the azimuthal modes with opposite rotation sense \cite{Zhu2005,Park2005,Neudecker2006,Ivanov2005,Zaspel2005,Hoffmann2007,Guslienko2008101,Guslienko2010}. The possibility that spin waves can reverse the vortex core has been predicted by analytical calculations and micromagnetic simulations \cite{Gaididei1999,Gaididei2000,Kovalev2002,Zagorodny2003,Kravchuk2007}. The first experimental observation of spin wave mediated vortex core reversal was published recently by applying bursts of relatively long rotating magnetic fields with a duration of $24$ periods \cite{Kammerer2011}. Experimental phase diagrams (i.e. excitation amplitude vs. frequency) have been shown which differ significantly for CW and CCW rotating spin waves, in good agreement with micromagnetic simulations which reveal significant differences in the magnetization dynamics for CW or CCW rotating spin wave modes in the vicinity of the vortex core. This allows unidirectional switching of the vortex core. It could be demonstrated \cite{Kammerer2011} that this asymmetry in switching the vortex core to its up or down polarisation is caused by the gyrofield which have been introduced in \cite{Thiele1973,Guslienko2008101} 

In the present paper we report on experiments and micromagnetic simulations in order to speed up vortex core reversal by applying much shorter rotating GHz magnetic field bursts of about one period duration only (corresponding to, e.g., $222\, ps$ excitation time at $4.5\, GHz$). In that way unidirectional vortex core reversal (i.e., selective switching only for vortex core up or vortex core down respectively) could be achieved with switching times down to almost $200\, ps$. However, we will demonstrate by micromagnetic simulations that - depending on the sample geometry - a lower limit exists for the switching time of spin wave mediated unidirectional and single vortex core reversal. 

Our experiments have been carried out at the MAXYMUS scanning transmission X-ray microscope at BESSY II, Berlin, \cite{Follath2010}. Time-resolved imaging is achieved by a stroboscopic pump-and-probe technique where the samples are excited periodically, synchronized to the electron bunches of the storage ring \cite{JAESCHKE1992}. Time-resolved imaging is done by varying the time delay between the excitation (pump) and the imaging X-ray flashes (probe). In order to improve the signal to noise ratio of time-resolved measurements significantly by getting rid of low frequency fluctuations, a complete time scan is done at each pixel before moving to the next one. The MAXYMUS scanning X-ray microscope combines a lateral resolution of typically $25\, nm$, given by the Fresnel zone plate used, with a temporal resolution of about $35\, ps$, determined by the inherent time structure of the BESSY II storage ring in multi-bunch operation. 

The sample stack, prepared on a $100\, nm$ thick $Si_3N_4$ membrane consists of a $170\, nm$ thick cross-like copper stripline, suitable for linear and rotating magnetic field generation. CCW and CW rotation senses were achieved by superposing x and y magnetic field components with a phase shift of $\pm 90$ degree. The previous experimental set-up and stripline design \cite{Curcic2011} had to be modified significantly for much higher frequencies up to $12\, GHz$. On top of the copper stripline a Permalloy disc with a diameter of $1.6\, \mu m$ and a thickness of $50\, nm$ were prepared by electron beam lithography and lift-off patterning.

We explored the switching behavior of the vortex core by applying short magnetic field bursts, each with a one-period duration and at an excitation frequency of $4.5\, GHz$ as sketched in Fig.~\ref{Fig:FastRotFieldsSwitching}, row B. For selective unidirectional switching of the vortex core polarity from down to up a burst with a CW sense of rotation has to be applied, followed by a burst with CCW sense of rotation which switches the vortex core polarity from up to down again. In that way time-resolved stroboscopic imaging was done in time steps as low as $12\, ps$. Fig.~\ref{Fig:FastRotFieldsSwitching} (A) shows selected time frames in intervals of $58\, ps$ displaying the out-of-plane magnetization of the spin waves during the CW excitation. The resulting vortex core up is again reversed, now by applying a magnetic field burst with CCW sense of rotation and the corresponding snapshots are shown in row C of Fig.~\ref{Fig:FastRotFieldsSwitching}. 

\begin{figure}
	\centerline{\includegraphics[clip,width=0.5\textwidth]{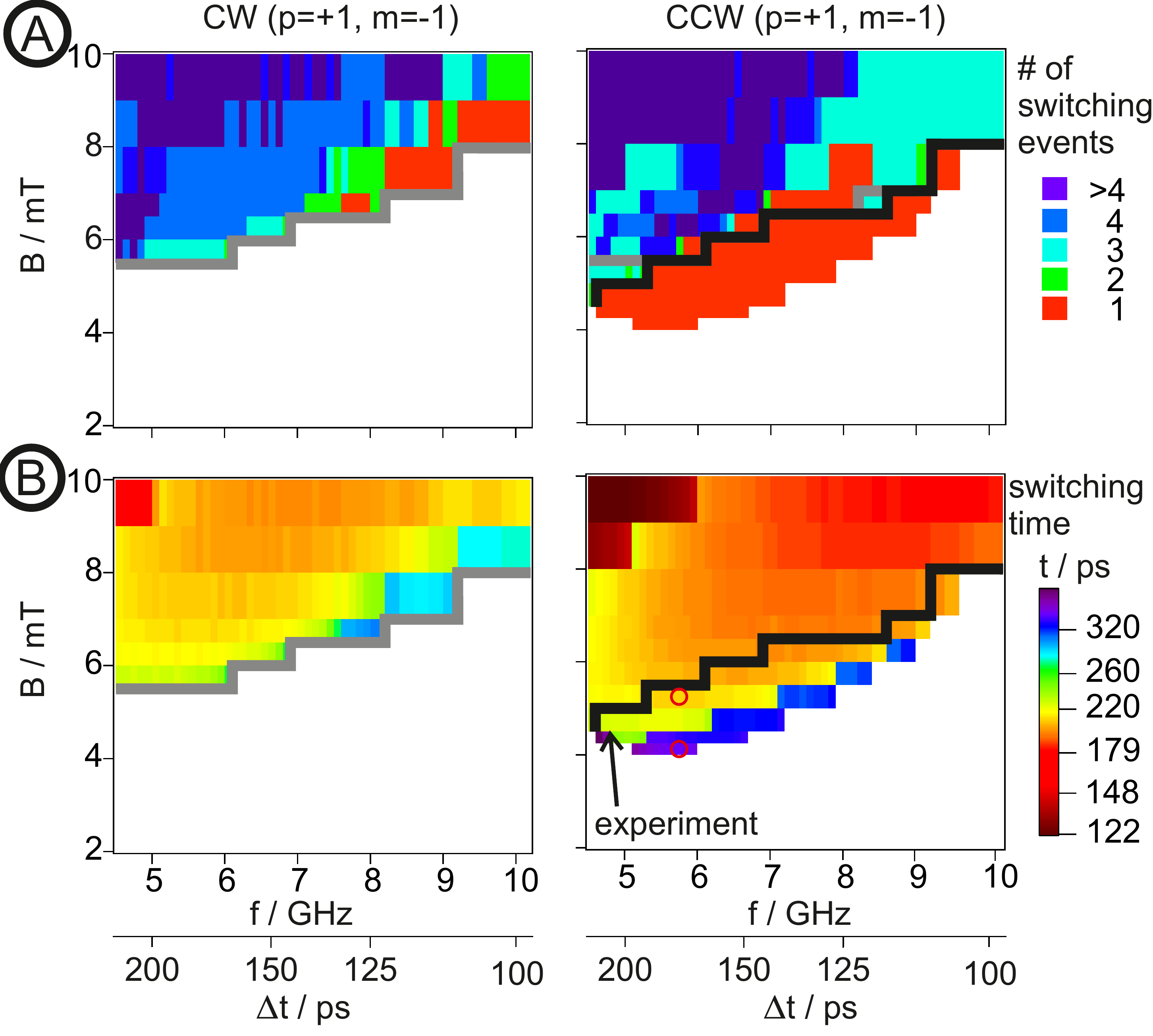}}
	\caption{{\bf Phase diagrams (amplitude vs. frequency) for vortex core switching by a one-period burst of a rotating magnetic ac excitation field starting with vortex core up.}
		Left column: CW excitation. Right column: CCW excitation. A (top): Number of switching events. B (bottom): Switching times. For details see text. }
		\label{Fig:Period1SwitchingGr_2}
\end{figure}

The clear contrast in all stroboscopic images shown in Fig.~\ref{Fig:FastRotFieldsSwitching} proves that for the parameter set used in our experiment (i.e., one-period burst at $4.5\, GHz$ with an excitation field amplitude of $4\, mT$) an unidirectional (selective) switching is achieved, to vortex core up by a CW and to vortex core down by a CCW burst excitation - as otherwise the contrast of the vortex core polarity between each burst would be lost as a superposition of both states would occur. Later in this paper it will be shown, that this experimental finding is also in good agreement with micromagnetic simulations predicting unidirectional switching for the chosen excitation parameters (cf. Fig.~\ref{Fig:Period1SwitchingGr_2}, B left). 

Experimental details on the developments of vortex magnetization of the vortex structure with vortex core up in the time region between $20\, ps$ and $70\, ps$ after CCW excitation has started are shown in Fig.~\ref{Fig:FastRotFieldsSwitching} D which can be compared with corresponding micromagnetic simulations in Fig.~\ref{Fig:FastRotFieldsSwitching} E. Magnetization profiles taken from the last time frames, $70.2\, ps$ after start of excitation, are printed in the last grey-shaded columns for both, the experimental and the simulated data. Both profiles are very similar and show, from left to right: 
(i) the magnetization of the positive part of the spin wave, (ii) the negative magnetization of the dip (i.e., the first stage of the developing vortex-antivortex pair which causes the reversal of the vortex core \cite{Kammerer2011}), (iii) the positive magnetization of the original vortex core, and (iv) the magnetization of the negative part of the spin wave. It should be noted that this configuration is found for vortex core switching by CCW (CW) rotating azimuthal spin waves and vortex core up (down), whereas for the two other configurations (i.e, CW for vortex core up and CCW for vortex core down) a different, more simple configuration is observed. This asymmetry was already observed by simulations in (\cite{Kammerer2011}, Fig.~4) and is caused by the gyrotropic field (\cite{Kammerer2011}, Fig.~5). More experimental evidence on this phenomenon will be given in the next paragraph.

\begin{figure}
	\centerline{\includegraphics[clip,width=0.5\textwidth]{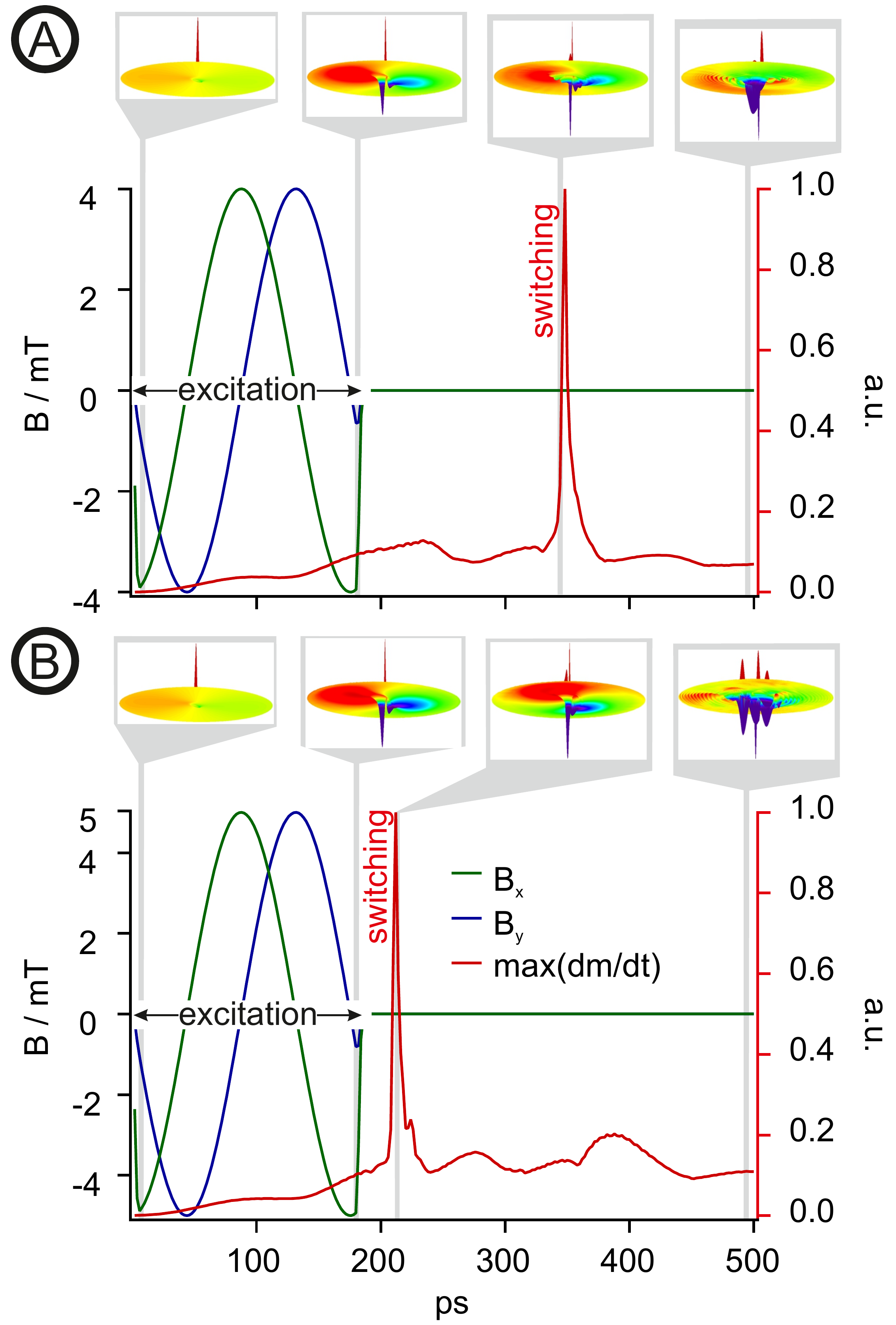}}
	\caption{{\bf Micromagnetic simulations demonstrating the delayed vortex core switching as a function of excitation amplitude.}
		Green / blue curve: maximum temporal derivation indicating the vortex core reversal. A (top graph): excitation amplitude: $4 \, mT$, duration of excitation: $175\, ps$. The vortex core switches after $342\, ps$ ($163\, ps$ after the end of the excitation). B (bottom graph): The same duration of the excitation, but with an amplitude of $5\, mT$. The vortex	core switches after $210\, ps$ ($35\, ps$ after the end of the excitation). The insets show the colour coded perpendicular magnetization of the vortex structure at times indicated by the gray lines. 
		A further increase of the excitation amplitude above $5\, mT$ is not favourable as it results in multiple vortex core switching. 
	}
	\label{Fig:SwitchingAfterBurst5700MHz4000muTRun933_02}
\end{figure}

\begin{figure}
	\centerline{\includegraphics[clip,width=0.5\textwidth]{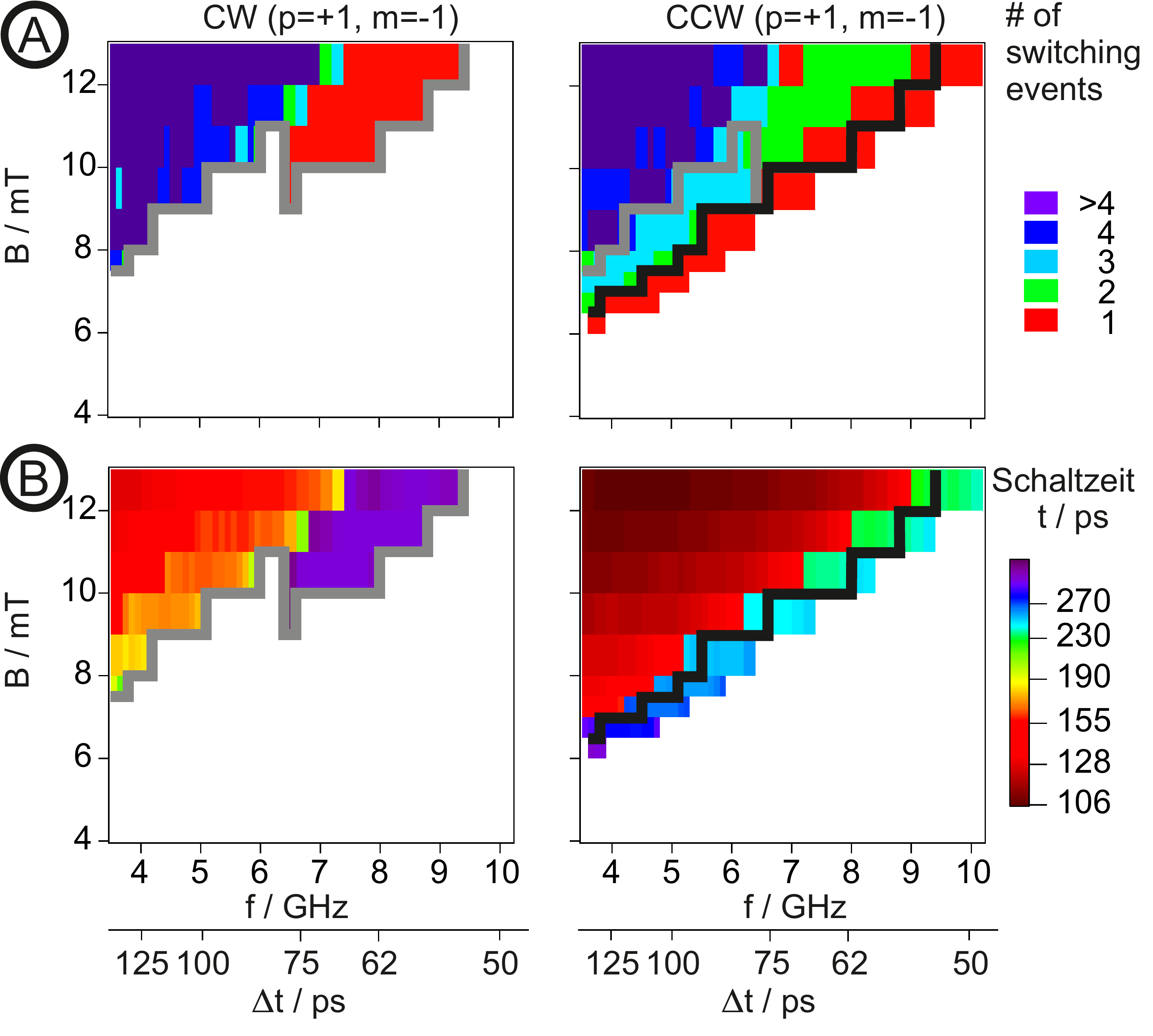}}
	\caption{{\bf Phase diagrams (amplitude vs. frequency) for vortex core switching by half a period burst excitation with a rotating magnetic ac excitation field starting with a vortex core up.}
		Left column: CW excitation. Right column: CCW excitation. A (top): Number of switching events. B (bottom): Switching times. For details see text. 	
	}
	\label{Fig:Period12SwitchingGr}
\end{figure}

In order to study in more detail the asymmetry in switching the vortex core by azimuthal spin waves of opposite sense of rotation we excited the system by continuously rotating fields with a frequency of $5.0\, GHz$ for the CW rotating spin wave mode ($n=1, m=-1$) and $5.5\, GHz$ for the CCW rotating mode ($n=1, m=+1$). The excitation amplitudes, $0.3\, mT$ for CW excitation and $0.9\, mT$ for CCW excitation, have been chosen just below the critical experimental value for core reversal at these specific frequencies \cite{Kammerer2011}. In contrast to the previous experiment (Fig.~\ref{Fig:FastRotFieldsSwitching}) where the magnetization configuration to be investigated could be observed only in a very short time frame before vortex core reversal, the continuous excitation just below the switching threshold level allows us to improve statistical accuracy significantly by measuring this magnetization configuration in a stationary state over a much longer time. For both, CW and CCW excitation, snapshots of the rotating magnetization have been taken at $25$ different phases of the rotation. These snapshots have been corrected in their rotation phase before averaging. The results are plotted in Fig.~\ref{Fig:RotationSum} A (left CW ($m=-1$), right CCW ($m=+1$)) and compared with the corresponding micromagnetic simulations, folded with the experimental resolutions (Fig.~\ref{Fig:RotationSum} B). In the case of CW excitation (left column), both, experiment and simulation reveal, that the vortex core (black circle) as well as the negatively polarized dip (white circle) is hidden behind the dipolar spin wave structure due to the limited lateral resolution. However, for CCW excitation, the vortex core (black circle) and the dip (white circle) exhibit a clear contrast due to their opposite polarization with respect to the spin wave profile. This agrees well with the snapshots and profiles in Fig.~\ref{Fig:FastRotFieldsSwitching} (row C for experiments, row D for simulations). This asymmetry, which has already been described by simulations before (\cite{Kammerer2011,Guslienko2010}), could now be imaged for the first time by an experiment. It is this asymmetry which causes the differences in the phase diagrams (Fig.~\ref{Fig:Period1SwitchingGr_2} and \ref{Fig:Period12SwitchingGr}) for CW or CCW rotating spin waves and therefore allows unidirectional vortex core switching with very short broad-band bursts as described in the following paragraph.

We have performed micromagnetic simulations for a Permalloy disc according to the measured sample size (diameter: $1.6\,\mu m$, thickness: $50\,nm$) in order to explore vortex core switching for a single-period excitation as a function of excitation frequency and field amplitude. The initial state of the vortex core was up ($p=+1$) in each of the simulations. The resulting phase diagrams are shown in Fig.~\ref{Fig:Period1SwitchingGr_2}. The left column denotes CW, the right column CCW excitation. (For starting with vortex core down ($p = -1$) just interchange CW and CCW.) In the upper row (A) the number of switches during and after one burst is given as a function of excitation frequency and amplitude\footnote{Please note, that because of the high-frequency bandwidth of an excitation with one period only, narrow resonance frequencies are no longer observed, which are present for a $24$-period excitation (\cite{Kammerer2011}, Fig.~2)}.

For reliable technological applications only the areas of single vortex core switching are relevant, which are marked red. Unidirectional switching is achieved at those frequencies and amplitudes, where the vortex core can be reversed with one sense of rotation only. Both conditions, unidirectional as well as single vortex core switching, are fulfilled for CCW excitation (cf. the upper right phase diagram in Fig.~\ref{Fig:Period1SwitchingGr_2}) in a wide area below the black line. Thus excitation frequencies from $4.5$ to $9.5\, GHz$ and excitation amplitudes between $4$ and $7\, mT$ (depending on the excitation frequency) can be applied for unidirectional and single vortex core reversal with one period of the excitation frequency. 

The next important information, which is derived from our simulations are the switching times which are shown in Fig.~\ref{Fig:Period1SwitchingGr_2} B. In the area of unidirectional and single vortex core reversal, switching times as low as $200\, ps$ are obtained. In our experiment shown in Fig.~\ref{Fig:FastRotFieldsSwitching} we obtained unidirectional vortex core reversal by a one-period excitation at $4.5\, GHz$ (corresponding to an excitation time of $220\, ps$) at an amplitude of $4.5\, mT$. At the corresponding point in the phase diagram of our simulations (marked 'experiment' in Fig.~\ref{Fig:Period1SwitchingGr_2}, B right) a small window for unidirectional, single vortex core reversal around $4.5\, mT$ is found, which is in sufficient agreement with our experiment. According to the simulations, the switching time would be around $300\, ps$, which corresponds to the sixth frame in Fig.~\ref{Fig:FastRotFieldsSwitching} A and \ref{Fig:FastRotFieldsSwitching} C. Unfortunately, the ongoing rotation of the spin waves prevents us to derive a precise vortex core reversal time from the experimental data.

An important finding is, that the time needed for vortex core reversal exceeds the excitation time. In other words, a delay has been discovered between the end of excitation and the actual vortex core reversal event. To give an example: One period at $8\, GHz$ corresponds to a $125\, ps$ excitation time, but at $5.5\, mT$ excitation amplitude vortex core reversal takes longer than $300\, ps$ (marked blue in Fig.~\ref{Fig:Period1SwitchingGr_2}, lower right diagram). Also in some other areas the switching time exceeds the excitation times by up to several $100\, ps$. Two more examples which are marked at $5.7\, GHz$ by the red circles in the right diagram of Fig.~\ref{Fig:Period1SwitchingGr_2} B will be treated in more detail in Fig.~\ref{Fig:SwitchingAfterBurst5700MHz4000muTRun933_02} A and B.

The delayed vortex core reversal is further investigated by more detailed analysis of micromagnetic simulations shown in Fig.~\ref{Fig:SwitchingAfterBurst5700MHz4000muTRun933_02}. The same vortex structure as treated before was excited by a rotatiting one-period field at $5.7\, GHz$, but at two different amplitudes, 4 mT in the upper (A) and $5\, mT$ in the lower (B) image. The green and blue curve denotes the excitation fields in x and y direction (left scale), the red curve, corresponding to the left scale, shows the maximum temporal derivative of the magnetization. It was found, that the latter is a perfect indicator for vortex core reversal as it exhibits a sharp peak exactly when the original vortex core is annihilated.

At the same excitation time of $(5.7\,GHz)^{-1 } = 175 \, ps$ the vortex core reversal time significantly decreased from $342$ (Fig.~\ref{Fig:SwitchingAfterBurst5700MHz4000muTRun933_02} A) to $210\, ps$ (Fig.~\ref{Fig:SwitchingAfterBurst5700MHz4000muTRun933_02} B) when the excitation amplitude is increased from $4$ to $5\, mT$. However, a further increase of the amplitude results in multiple switching, where selectivity cannot be assured anymore. Following the phase diagram (Fig.~\ref{Fig:Period1SwitchingGr_2} B, right plot), the single switching time for this sample geometry at a one-period excitation is limited to about $200\, ps$ at all excitation frequencies simulated ($4.5$ to $10\, GHz$). 

\begin{figure*}
	\centerline{\includegraphics[clip,width=\textwidth]{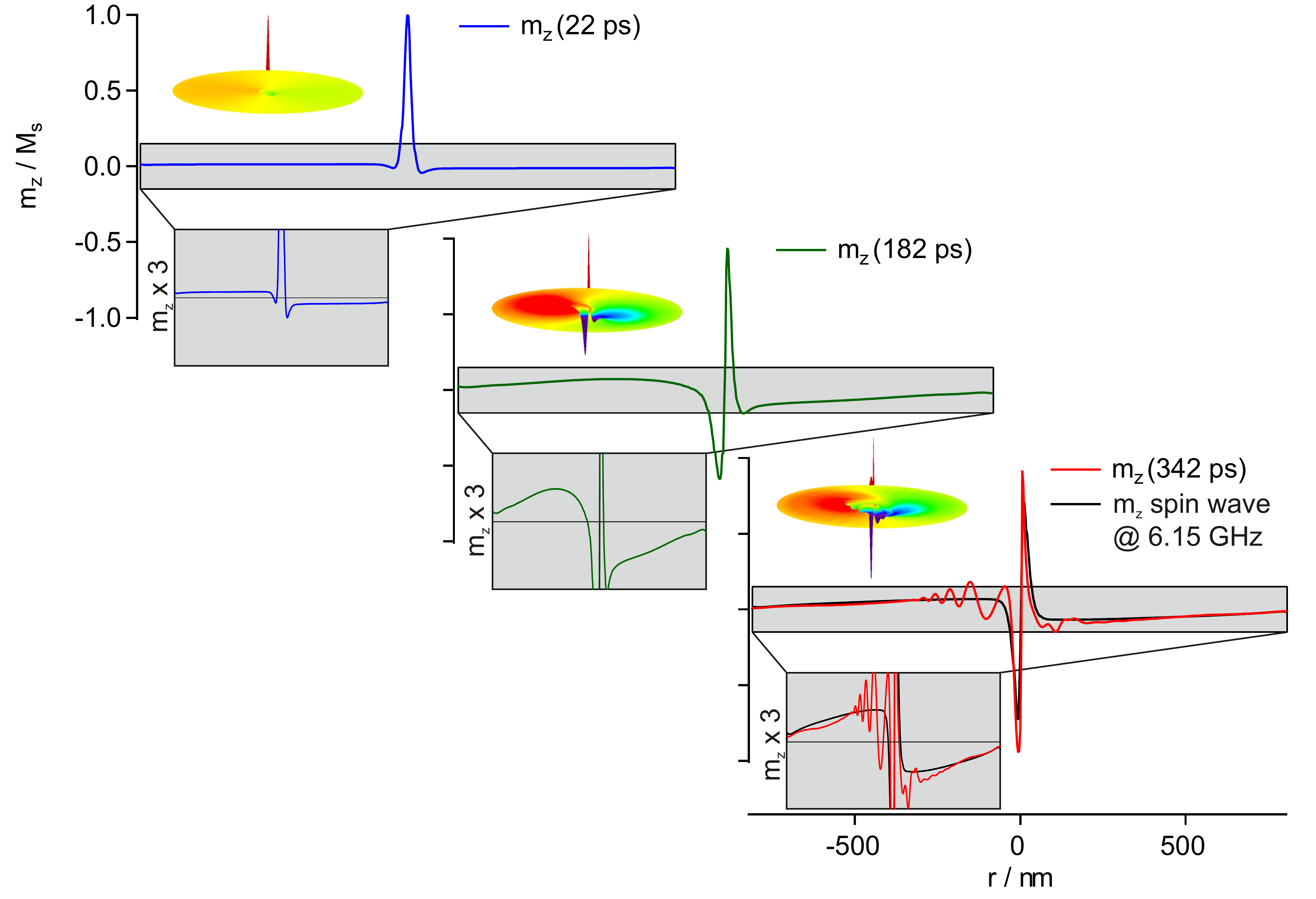}}
	\caption{{\bf Explanation of the delay in vortex core reversal by transport of energy towards the centre of the sample}.
		Cuts through the simulated evolution of the magnetization in z direction are plotted for three time steps during the same burst excitation as applied for the top graph in Fig.~\ref{Fig:SwitchingAfterBurst5700MHz4000muTRun933_02} (one-period excitation at $5.7\, GHz$, field amplitude: $4\, mT$). The insets to the bottom left show an extension of the z-axis for the inner region of interest.
		Left ($22\, ps$ after excitation has started): a radially homogeneous precession is uniformly excited.
		Middle ($182\, ps$ after excitation has started and shortly after excitation has stopped (cf. Fig.4): A concentration of the magnetization amplitude towards the centre of the disc is observed.
		Right ($342\, ps$ after excitation has started and $167\, ps$ after excitation has stopped): The enhanced concentration of the magnetization in the centre leads to the formation of the negative dip, followed by the formation of a vortex-antivortex pair and vortex core reversal. For comparison, the magnetization of the excited spin wave eigenmode is also shown by the black line in this graph.
	}
	\label{Fig:SpinWaveTransformationReal}
\end{figure*}

In a next step we have investigated by micromagnetic simulations if the above limit in vortex core reversal time might be overcome by a shorter excitation time. For that purpose we excited the same sample with a reduced excitation time of half a period (c.f. Fig.~\ref{Fig:Period12SwitchingGr}). Comparison with a one-period excitation (Fig.~\ref{Fig:Period1SwitchingGr_2}) shows that also for half a periode excitation single switching can only be achieved with CCW excitation. compared to the one-period excitation, a higher switching threshold is observed in Fig.~\ref{Fig:Period12SwitchingGr} A, increased by about a factor of $\sqrt{2}$ since an approximately constant energy for switching is assumed which scales (in a linear approximation) with $B^2\cdot t$. The most important finding is that for half a period excitation the switching time for single switching is again found to be in the order of $200-300\, ps$. Thus the limitation of the vortex core switching time of about $200\, ps$ according to Figs.~\ref{Fig:Period1SwitchingGr_2} B and Fig.~\ref{Fig:SwitchingAfterBurst5700MHz4000muTRun933_02} B can not be overcome by the shorter excitation time. 

We would like to suggest an explanation for the limitation of the vortex core switching time discovered in our simulations. This explanation is based on a finite time for the energy transport (with spin wave velocities) from the outer area of the sample to a location near to the vortex core. Roughly speaking, the radially homogeneous excited spin wave amplitudes are concentrated in the center due to the concentric shape of the disc geometry. 
This is supported by additional micromagnetic simulations on the development of the normalized magnetization $m_z$ in z direction as a function of time as sketched in Fig.~\ref{Fig:SpinWaveTransformationReal}. The excitation parameters are the same as for Fig.~\ref{Fig:SwitchingAfterBurst5700MHz4000muTRun933_02} A (excitation amplitude: $4\, mT$). At three distinct time steps (please compare with the time scale in Fig.~\ref{Fig:SwitchingAfterBurst5700MHz4000muTRun933_02} A) one dimensional profiles are plotted of the magnetization at the symmetry axis of the spin wave. In the left plots in Fig.~\ref{Fig:SpinWaveTransformationReal} the blue line denotes the radially homogeneous out-of-plane magnetization attributed to a precessional motion at the beginning of the excitation ($22\, ps$ after the excitation has started) The inlay shows the same magnetization, by with a field scale enhanced by a factor of three. In the mid plots of Fig.~\ref{Fig:SpinWaveTransformationReal}, $163\, ps$ later and at the end of the excitation time of $175\, ps$, the magnetization amplitude is already more concentrated to the middle of the sample as shown by the green profile. And finally in the right plots at $342\, ps$, the concentration of the magnetization near the vortex core (red line) is high enough to form the vortex-antivortex-pair causing the vortex core reversal $163\, ps$ after excitation has ended. For comparison, the black line denotes the magnetization of the CCW rotating spin wave mode ($n = 1$, $m = -1$) with an eigenfrequency of about $6.15\, GHz$. 
Acording to a rough estimation, the velocity of the energy transport to the center is approximately in the order of $1000\, m/s$ and thus matches with the typical velocities of magnetostatic spin waves. 

Our simulations show that transport of the global energy towards the center of the vortex structure has to be considered for vortex core switching. In our sample ($1.6\, \mu m$ in diameter, $50\, nm$ thick) the energy transport limits the switching time for singular vortex core reversal to about $200\, ps$, even for shorter excitation bursts. However, as energy transport time (for a rough estimation) is in first order proportional to the sample size, our model implies that much shorter vortex core reversal times can be obtained at smaller samples. 

To conclude, we have experimentally imaged spin wave mediated unidirectional vortex core reversal by time-resolved X-ray microscopy. The vortex core in a Permalloy disc, $1.6\,\mu m$ in diameter and $50\, nm$ thick, could be switched experimentally by a one-period ($222\,ps$) rotating ac magnetic field burst at $4.5\, GHz$. Unidirectional switching to up or down could be achieved by adjusting the sense of rotation of the ac field burst, in good agreement with the phase diagrams obtained by micromagnetic simulations. Simulations also show that this unidirectional (selective) vortex core reversal is possible due to significant differences for CW or CCW rotating azimuthal spin waves, based on the interaction of the gyrofield \cite{Thiele1973,Guslienko2008100} with the spin wave \cite{Kammerer2011}. These differences in vortex magnetization dynamics during excitation of CW or CCW rotation spin waves could now be imaged in an experiments by time-resolved scanning transmission X-ray microscopy. 

Significant delays between the end of the excitation and vortex core reversal time of up to several $100\, ps$ have been found by micromagnetic simulations, depending on excitation frequency and amplitudes. This delay time can be reduced by increasing the excitation amplitude which is, on the other hand, limited due to multiple switching. As a consequence a lower limit for unidirectional vortex core reversal is observed (about $200\, ps$ for a sample geometry of $1.6\, \mu m$ in diameter and a thickness of $50\, nm$). This phenomenon is explained by the time needed for the energy transport of the global excitation towards the center.

\begin{acknowledgments}
We would like to thank Manfred Fähnle, Eberhard Göring and Dieter Carstanjen, MPI Stuttgart for fruitful discussions. 
\end{acknowledgments}

 \end{document}